\documentclass{pasj00}

\begin{document}
\SetRunningHead{H. Yamaguchi et al.}{Fe-K Emission in RCW~86 Northeast}
\Received{2007/08/24}
\Accepted{2007/09/19}

\title{Suzaku Observation of the RCW~86 Northeastern Shell}

\author{Hiroya \textsc{Yamaguchi} and Katsuji \textsc{Koyama}}
\affil{Department of Physics, Kyoto University, 
  Kitashirakawa-oiwake-cho, Sakyo-ku, Kyoto 606-8502}
\email{hiroya@cr.scphys.kyoto-u.ac.jp}

\author{Hiroshi \textsc{Nakajima}}
\affil{Department of Earth and Space Science, Osaka University, 
  1-1 Machikaneyama, Toyonaka, Osaka 560-0043}

\author{Aya \textsc{Bamba}}
\affil{Institute of Space and Astronautical Science, JAXA, 
  3-1-1 Yoshinodai, Sagamihara, Kanagawa 229-8510}

\author{Ryo \textsc{Yamazaki}}
\affil{Department of Physics, Hiroshima University, 
  Higashi-Hiroshima, Hiroshima 739-8526}

\author{Jacco \textsc{Vink}}
\affil{Astronomical Institute, University Utrecht, 
P.O. Box~80000, 3508TA Utrecht, Netherlands}

\author{and \\ Akiko \textsc{Kawachi}}
\affil{Department of Physics, Tokai University, 
Kitakaname 1117, Hiratsuka, Kanagawa 259-1292}

\maketitle

\begin{abstract}

This paper reports on Suzaku results concerning the northeast shell of RCW~86. 
With both spatial and spectral analyses, 
we separated the X-rays into three distinct components: 
low ($kT_e\sim$0.3~keV) and high ($kT_e\sim$1.8~keV) temperature plasmas 
and a non-thermal component, and discovered that 
their spatial distributions are different from each other. 
The low-temperature plasma is dominated at the east rim, whereas  
the non-thermal emission is brightest at the northeast rim, 
which is spatially connected from the east. 
The high-temperature plasma, found to contain the $\sim$6.42~keV line 
(K$\alpha$ of low-ionized iron), is enhanced at the inward region 
with respect to the east rim, and has no spatial correlation with 
the non-thermal X-ray (the northeast). 
This result suggests that the Fe-K$\alpha$ line originates from 
Fe-rich ejecta heated by reverse shock. 
A possible scenario to explain these morphologies and spectra is that 
a fast-moving blast wave in a thin cavity collided with 
a dense interstellar medium at the east region very recently. 
As a result, the reverse shock in this interior decelerated, 
and arrived at the Fe-rich region of the ejecta and heated it. 
In the northeast rim, on the other hand, the blast wave is still moving fast, 
and is accelerating electrons causing them to emit strong synchrotron X-rays.

\end{abstract}

\section{Introduction}
\label{sec:introduction}

RCW~86 (G315.4--2.3) is one of the historical Galactic 
supernova remnants (SNRs), a possible remnant of supernova 
in AD~185 (Stephenson \& Green~2002). 
The distance to RCW~86 was determined to be 2.8~kpc by 
optical observations (Rosado et al.~1996); we assume this value here. 
The ASCA observations discovered synchrotron X-rays from the southwest (SW) 
and northeast (NE) shells of RCW~86 (Bamba et al.~2000; Borkowski et al.~2001), 
which indicates that the energy of the electrons reaches up to $\sim$100~TeV 
by shock acceleration, similar to the mechanism in SN~1006 
(Koyama et al.~1995).

Another remarkable discovery with ASCA is the $\sim$6.4~keV line 
in the spectra of the SW (Vink et al.~1997) and 
NE rims (Tomida et al.~1999) of RCW~86. 
Since the center energy of the line is consistent with K$\alpha$ 
from neutral Fe, a fluorescent origin had often been proposed 
(Vink et al.~1997; Tomida et al.~1999). 
However, Bamba et al.~(2000) and Borkowski et al.~(2001) showed that 
the X-ray spectrum of the SW shell was could be represented by 
a three-component model; two thin-thermal plasmas with 
low ($\sim$0.3--0.8~keV) and high ($>$5~keV) temperatures, 
and a non-thermal component. 
The Fe-K$\alpha$ line was explained by the high-temperature plasma with an
extremely low ionization parameter ($\tau = n_et$) of $\sim 10^9$~cm$^{-3}$~s.

Rho et al.~(2002) observed the SW shell with Chandra, and 
spatially resolved the low-temperature plasma and non-thermal emission. 
Since the low-temperature plasma is spatially correlated with 
optical H$\alpha$ emission, they concluded the origin to be a blast wave. 
On the other hand, the non-thermal emission is localized 
at the inner region of the low-temperature plasma; 
hence, the origin was suggested to be reverse shock. 
They found, furthermore, that the Fe-K$\alpha$ correlated with 
the non-thermal emission, and suggested that the Fe-K$\alpha$ line 
originates from a high-temperature plasma of Fe-rich ejecta heated 
by the synchrotron-emitting reverse shock. 
However, the morphology of the Fe-K$\alpha$ emitting region 
could not be determined.

\begin{figure*}[!htb]
  \begin{center}
    \FigureFile(80mm,80mm){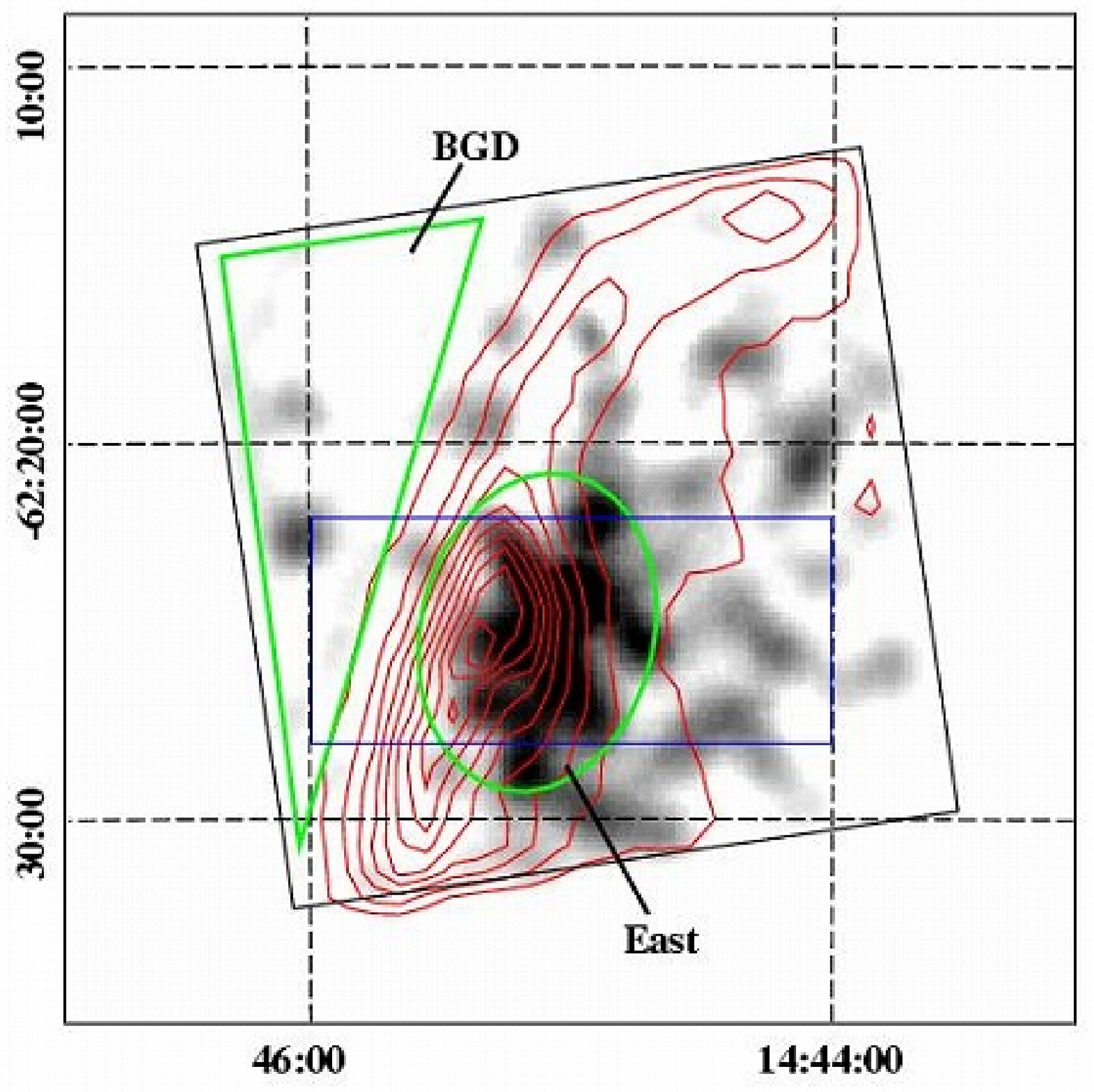}
    \FigureFile(80mm,80mm){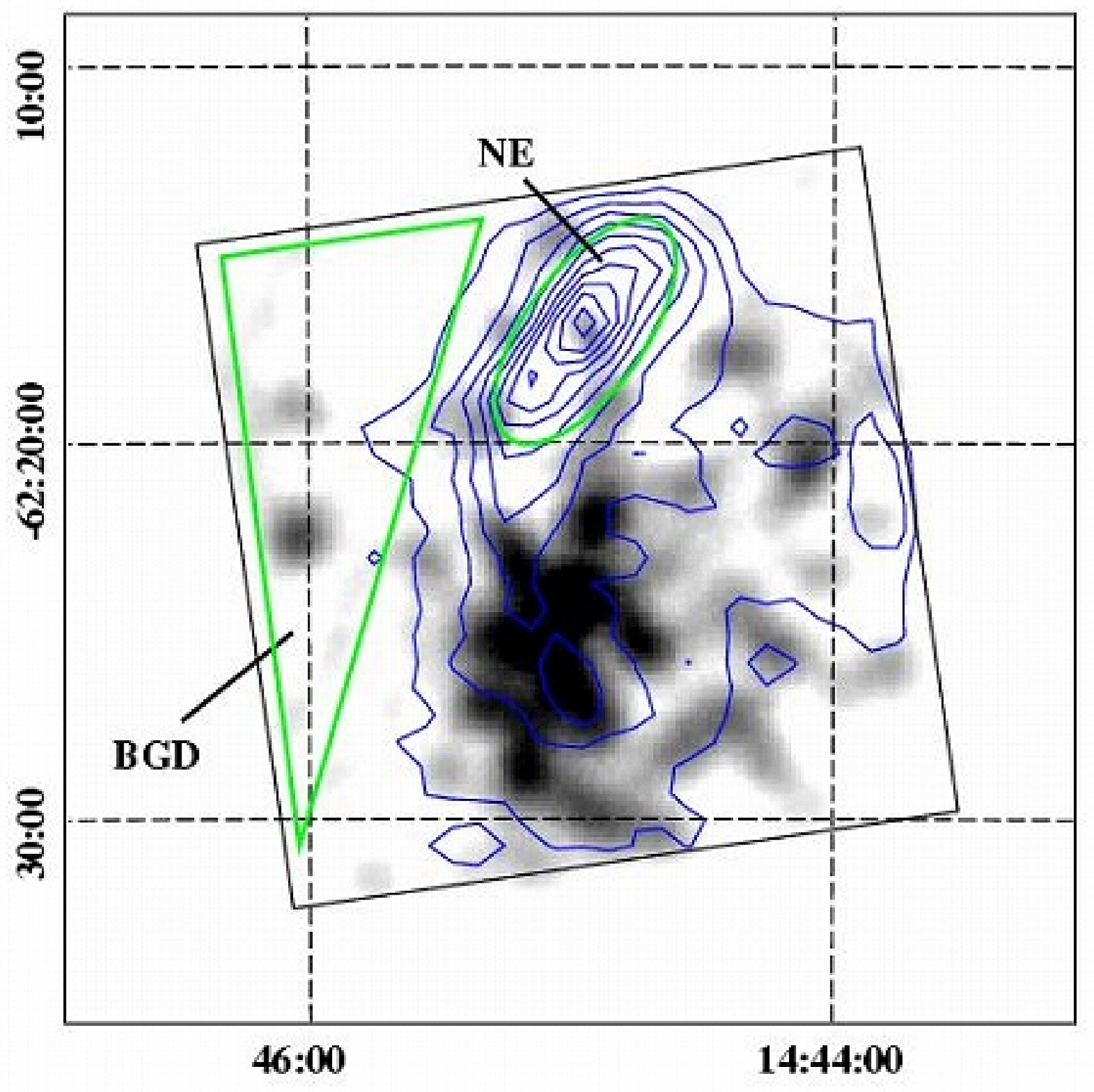}
  \end{center}
  \caption{0.5--1.0~keV (left) and 3.0--6.0~keV (right) intensity contours 
    overlaid on the Fe-K map (gray scale) smoothed with 
    a Gaussian kernel of $\sigma = 1.1$~arcmin. 
    The coordinates (RA and Dec) refer to epoch J2000.0. 
    The Field of view of the XIS is indicated by the black squares, 
    where the data from the four XIS are combined, but that of 
    the four corners, irradiated by the $^{55}$Fe calibration sources, 
    are removed. 
    The scales of contours and gray scale are linear, and these ranges are 
    from 0 to $6.3\times 10^{-2}$~counts~s$^{-1}$~arcmin$^{-2}$ 
    (red contour), 
    to $1.6\times 10^{-2}$~counts~s$^{-1}$~arcmin$^{-2}$ (blue contour), 
    and to $2.0\times 10^{-3}$~counts~s$^{-1}$~arcmin$^{-2}$ (gray scale). 
    The green ellipses and the triangles are the source 
    and background regions, respectively.
    The blue rectangle indicates the region where 
    the projection profiles are extracted (see figure~\ref{fig:proj}).}
  \label{fig:image}
\end{figure*}

Recently, the SW region of RCW~86 was observed with 
the X-ray Spectrometers (XIS; Koyama et al.~2007) 
aboard the new X-ray satellite Suzaku (Mitsuda et al.~2007). 
Utilizing the good sensitivity and spectroscopic performances of the XIS 
in the energy band of Fe-K lines for diffuse sources, 
Ueno et al.~(2007) determined the center energy of the Fe-K$\alpha$ line 
to be 6404~(6400--6407)~eV, which indicates the emission is 
from low-ionized iron less than Si-like (Fe\emissiontype{XIII}). 
Furthermore, they revealed the Fe-K line morphology for the first time 
and discovered its distribution is spatially different from 
that of the non-thermal emission. 
They thus confirmed that the origin of the Fe-K$\alpha$ is not fluorescence 
caused by supra-thermal electrons, nor non-thermal X-rays, 
and concluded that a more likely origin is 
a high-temperature ejecta in extremely ionization non-equilibrium.

The NE region of RCW~86 also exhibits Fe K-shell emission 
(albeit the flux is less than that of SW) as well as soft thermal and 
non-thermal X-rays (ASCA: Tomida et al.~1999). 
Therefore, this region would be another good place to solve 
the puzzle concerning the origin of Fe-K$\alpha$ related to 
the thermal and non-thermal emission in RCW~86. 
Chandra and XMM-Newton revealed that the soft-thermal and non-thermal 
emission filaments join smoothly along the outer shell (Vink et al.~2006). 
They, however, failed to detect Fe-K$\alpha$ from this region, 
possibly due to the limited sensitivity and high background level 
near the Fe K-shell energy. 
We therefore observed the NE shell with Suzaku 
to see if the Fe-K$\alpha$ is really present or not. 
If present, we will study its morphology and spectra to reveal the origin.

\section{Observations and Data Reduction}
\label{sec:observation}

The Suzaku observation of the NE region of RCW~86 was made on 2006 
August 12 (Observation ID = 501037010), using the four XIS placed on 
the focal plane of an X-Ray Telescope (XRT; Serlemitsos et al.~2007). 
The XIS consists of three Front-Illuminated (FI) CCDs 
and one Back-Illuminated (BI) CCD. 
The advantages of the former are high detection efficiency and 
low background level in the energy band above $\sim$5~keV, 
while the latter has significantly superior sensitivity in 
the energy band of $\lesssim$1~keV with moderate energy resolution. 
All four XRTs are co-aligned to image the same region of the sky.
The XIS were operated in the normal full-frame clocking with 
the $3\times 3$ or $5\times 5$ editing mode. 
We employed cleaned revision 1.2 data, and used the HEADAS software 
version 6.0.4 and XSPEC version 11.3.2 for the data reduction and analysis. 
After screening, the effective exposure time was obtained 
to be $\sim$53~ksec. 
The response matrix files (RMF) and ancillary response files (ARF) were 
made using xisrmfgen and xissimarfgen (Ishisaki et al.~2007) 
version 2006-10-17.

The errors quoted in the text and tables are at the 90\% confidence level, 
and the 1$\sigma$ confidence level in the figures, unless otherwise stated.

\section{Analysis and Results}
\label{sec:analysis}

\begin{table*}[!t]
  \begin{center}
    \caption{Best-fit parameters of 3--10~keV fittings.}
    \label{tab:hard}
    \begin{tabular}{llcc}
      \hline 
      Component  &  Parameter  &  East     &  NE       \\
      \hline
           Power-law  & $\Gamma$    & 3.18~(3.02--3.31)  & 2.96~(2.88--3.05) \\
      ~       & Flux$^{\ast}$  & 1.10~(1.07--1.14)~$\times 10^{-4}$  &
                                        2.44~(2.39--2.48)~$\times 10^{-4}$ \\
      ~ & Surface brightness$^{\ast}$ & 2.68~(2.61--2.78)~$\times 10^{-6}$  &
                                        1.36~(1.33--1.38)~$\times 10^{-5}$ \\
      Gaussian   & Center~(eV) & 6424~(6404--6444)  & 6424~(fixed)      \\
      ~  & Width$^{\dagger}$~(eV)  &  $<$90         & 0~(fixed)         \\
      ~  & Flux$^{\ddagger}$   & 7.1~(5.8--8.3)~$\times 10^{-6}$  &
                                                $<$8.1~$\times 10^{-7}$ \\
      ~  & Surface brightness$^{\ddagger}$ & 1.7~(1.4--2.0)~$\times 10^{-7}$ &
                                                $<$4.8~$\times 10^{-8}$ \\
      \hline
      \multicolumn{2}{c}{$\chi ^2$/d.o.f.}  &  146/192 = 0.76 
                                                    & 204/234 = 0.87    \\
      \hline
      \multicolumn{4}{l}{$^{\ast}$
      Photon flux (photons~cm$^{-2}$~s$^{-1}$) and surface brightness } \\
      \multicolumn{4}{l}{\ \ 
      (photons~cm$^{-2}$~s$^{-1}$~arcmin$^{-2}$) 
      in the 3--10~keV band.} \\
      \multicolumn{4}{l}{$^{\dagger}$
      One standard deviation (1$\sigma$).} \\
      \multicolumn{4}{l}{$^{\ddagger}$
      Total flux (photons~cm$^{-2}$~s$^{-1}$) 
      and surface brightness} \\
      \multicolumn{4}{l}{\ \ 
      (photons~cm$^{-2}$~s$^{-1}$~arcmin$^{-2}$) 
      in the line.} \\
    \end{tabular}
  \end{center}
\end{table*}

\begin{figure}[!b]
  \begin{center}
    \FigureFile(80mm,80mm){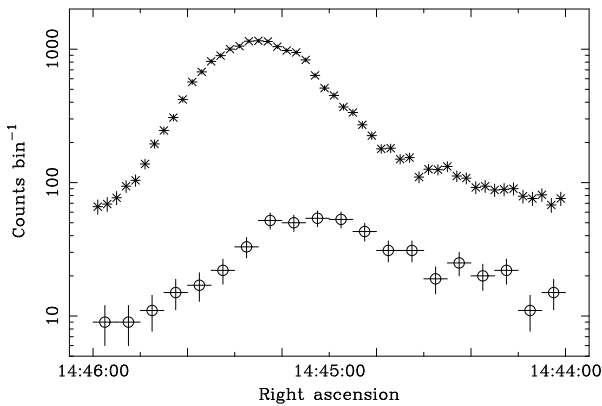}
  \end{center}
  \caption{Projection profiles (along Right ascension) of 
    the Ne\emissiontype{IX}-K$\alpha$ band (cross; 0.87--0.94~keV) and 
    the Fe-K$\alpha$ band (circle; 6.3--6.5~keV) 
    for the blue rectangular region given in figure~\ref{fig:image} left.
    The bin-size of the former corresponds to $\sim$\timeform{0.28'}, 
    while that of the latter is $\sim$\timeform{0.69'}. 
  }
  \label{fig:proj}
\end{figure}

\subsection{XIS Image}
\label{ssec:image}

We show the Fe-K$\alpha$ intensity map in figure~\ref{fig:image} in 
gray scale. This map was made from the 6.3--6.5~keV (the Fe-K$\alpha$ band) 
image by subtracting the continuum level in this band, 
which was estimated with a power-law model (with photon index of 3) 
fitting for the 5.0--6.2~keV continuum spectrum. 
For comparisons with the soft and hard X-ray fluxes, we overlaid 
intensity contours of the 0.5--1.0~keV band (figure~\ref{fig:image} left) 
and the 3.0--6.0~keV band (figure~\ref{fig:image} right). 
No spatial correlation between the Fe-K$\alpha$ and 
the hard X-ray emission was found. 
Moreover, the soft X-ray emission is systematically shifted 
outward with respect to the Fe-K$\alpha$ emission. 
As we show in subsection~\ref{ssec:full}, the 0.5--1.0~keV emission 
is dominated by an optically thin thermal plasma 
with an electron temperature of 
$\sim$0.3~keV, and the Ne\emissiontype{IX}-K$\alpha$ line is 
the major emission of this component. 
We then made projection profiles of the Ne\emissiontype{IX}-K$\alpha$ 
line band (0.87--0.94~keV), and the Fe-K$\alpha$ band (6.3--6.5~keV), 
in the position given in figure~\ref{fig:image} left. 
The result is shown in figure~\ref{fig:proj}. 
We can see that the peak of Fe-K$\alpha$ is significantly shifted to 
the inner region from the soft emission (Ne\emissiontype{IX}-K$\alpha$). 
With the fitting of a Gaussian-plus-constant model, 
the peak positions (right ascensions) of the soft plasma and 
Fe-K$\alpha$ are found to be separated by 2~arcmin.

\begin{figure}[!b]
  \begin{center}
    \FigureFile(80mm,80mm){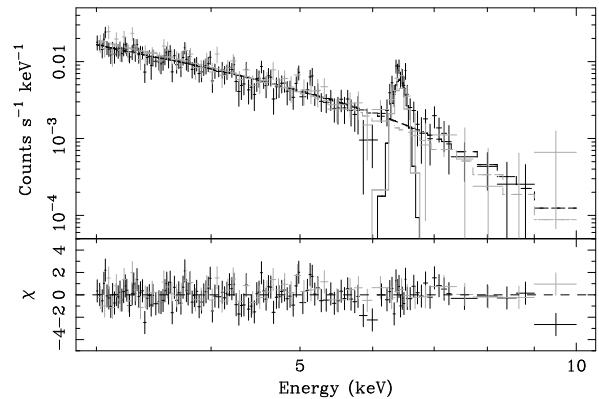}
  \end{center}
  \caption{XIS 3--10~keV spectra extracted from the East region. 
    Black and gray represent FI (sum of the three sensors) and BI, 
    respectively. 
    The best-fit power-law and Gaussian models are shown by solid lines. 
  }
  \label{fig:hard}
\end{figure}

\begin{table*}[htb]
  \begin{center}
    \caption{Best-fit parameters of full-band model fittings.}
    \label{tab:full}
    \begin{tabular}{llcc}
      \hline 
      Component   &  Parameter    &  East       &  NE         \\
      \hline
      Absorption  &  $N_{\rm H}$~(cm$^{-2}$) 
                                  &  4.1~(3.8--4.4)~$\times 10^{21}$
                                    &  4.4~(4.2--4.5)~$\times 10^{21}$     \\
      VPSHOCK~1   &  $kT_e$~(keV) &  0.33~(0.30--0.37) & 1.2~(0.89--1.6) \\
      ~           &  C, N, O      &  0.15~(0.11--0.24) & 0.15~(fixed)    \\
      ~           &  Ne           &  0.33~(0.26--0.50)  & 0.33~(fixed)       \\
      ~           &  Mg           &  0.31~(0.28--0.48)  & 0.31~(fixed)       \\
      ~           &  Si, S        &  0.31~(0.19--0.43)  & 0.31~(fixed)      \\
      ~           &  Ca, Fe, Ni   &  0.23~(0.17--0.36)  & 0.23~(fixed)   \\
      ~   & $\tau$~(cm$^{-3}$~s)  &  7.7~(5.5--11)~$\times 10^{10}$ 
                                       &  6.7~(5.9--7.7)~$\times 10^9$   \\
      ~   & $n_en_{\rm H}V$~(cm$^{-3}$) & 3.7~(2.4--5.8)~$\times 10^{57}$
                                    &  2.4~(2.0--2.9)~$\times 10^{56}$   \\
      VPSHOCK~2   &  $kT_e$~(keV) &  1.8~(1.4--2.3)  & --                \\
      ~           &  Fe           &  32~(14--90)     & --                \\
      ~   & $\tau$~(cm$^{-3}$~s)  &  2.3~(1.8--2.9)~$\times 10^9$  & --  \\
      ~   & $n_en_{\rm H}V$~(cm$^{-3}$) & 2.0~(1.1--2.6)~$\times 10^{56}$
                                                     & --                \\
      Power-law   & $\Gamma$   &  2.89~(2.45--3.28)  & 2.78~(2.74--2.82)  \\
      ~           & Norm$^{\ast}$ &  1.1~(0.72--1.9)~$\times 10^{-3}$
                                  &  3.3~(3.2--3.5)~$\times 10^{-3}$  \\
      \hline
      Gain~(FI)   & Offset~(eV)   & --3.2            & --3.2             \\
      Gain~(BI)   & Offset~(eV)   & --5.4            & --5.4             \\
      $\chi ^2$/d.o.f.  &~        & 1093/1050 = 1.04 & 1138/1084 = 1.05  \\
      \hline
      \multicolumn{4}{l}{$^{\ast}$
      The differential flux (photons~cm$^{-2}$~s$^{-1}$) at 1~keV.} \\
    \end{tabular}
  \end{center}
\end{table*}

\subsection{Spectrum in the Hard X-Ray Band}
\label{ssec:hard}

For a quantitative study, we extracted the representative XIS spectra 
from two elliptical regions: the East and NE (see figure~\ref{fig:image}), 
where the soft X-ray and Fe-K$\alpha$, 
and the hard X-ray flux are the brightest. 
We did not divide the former two diffuse emissions 
because the separation between them is comparable to the angular resolution 
of the XRT ($\sim$\timeform{2'}: Serlemitsos et al.~2007), and 
they are overlapping each other. 
The solid angles of these ellipses are 
$\pi \times \timeform{4.2'}\times \timeform{3.1'}$ = 41~arcmin$^2$ and 
$\pi \times \timeform{3.5'}\times \timeform{1.6'}$ = 18~arcmin$^2$, respectively. 
The background data were taken from the outside of the remnant 
(the solid triangles in figure~\ref{fig:image}). 
For the background spectra, we excluded point-like sources 
at (\timeform{14h46m03s},\timeform{-62D22'30"}) and
(\timeform{14h46m03s},\timeform{-62D19'01"}), and
the CCD corners, which contain the $^{55}$Fe calibration source.

We first determined the Fe-K$\alpha$ line parameters. 
The background-subtracted spectra in the 3--10~keV band 
in the East region is shown in figure~\ref{fig:hard}. 
Since the data from the three FIs are nearly identical, 
we merged the individual spectrum to improve the photon statistics.

We detected a clear Fe-K$\alpha$ line in the spectra. 
We then fitted the spectra with a power-law (for the continuum) 
and a Gaussian (for the emission line). 
The best-fit parameters are given in table~\ref{tab:hard}. 
The centroid energy of the Fe-K$\alpha$ line is slightly higher than 
that of the emission line from neutral iron (6400~eV). 
Although the absolute energy calibration error is reported to be 
as accurate as $\pm$0.2\%, above 1~keV (Koyama et al.~2007), we checked 
the energy scale of each XIS sensor using the $^{55}$Fe calibration sources. 
We then confirmed that the peak energies of Mn-K$\alpha$ lines were 
consistent with the laboratory value of 5895~eV 
within the statistical errors for all of the XIS sensors. 
We therefore conclude that the energy scale error 
in the Fe-K line band is less than the statistical error. 
The best-fit energy of 6424~(6404--6444)~eV constrains 
the Fe ionization state to be approximately between Ar-like and Ne-like.

For a comparison, we extracted the spectra from the NE region in the same way 
as that of the East, and fitted the spectra with a power-law plus a Gaussian. 
Since we could see no clear the Fe-K$\alpha$ line, we fixed the center of 
the Gaussian to be 6424~eV, and determined the upper limit of the line flux. 
The result is given in table~\ref{tab:hard}. 
The surface brightness of the Fe-K$\alpha$ and the power-law component 
in the NE are at least $\sim$3-times smaller and 
$\sim$5-times higher than those in the East region.

\begin{figure*}[htb]
  \begin{center}
    \FigureFile(80mm,80mm){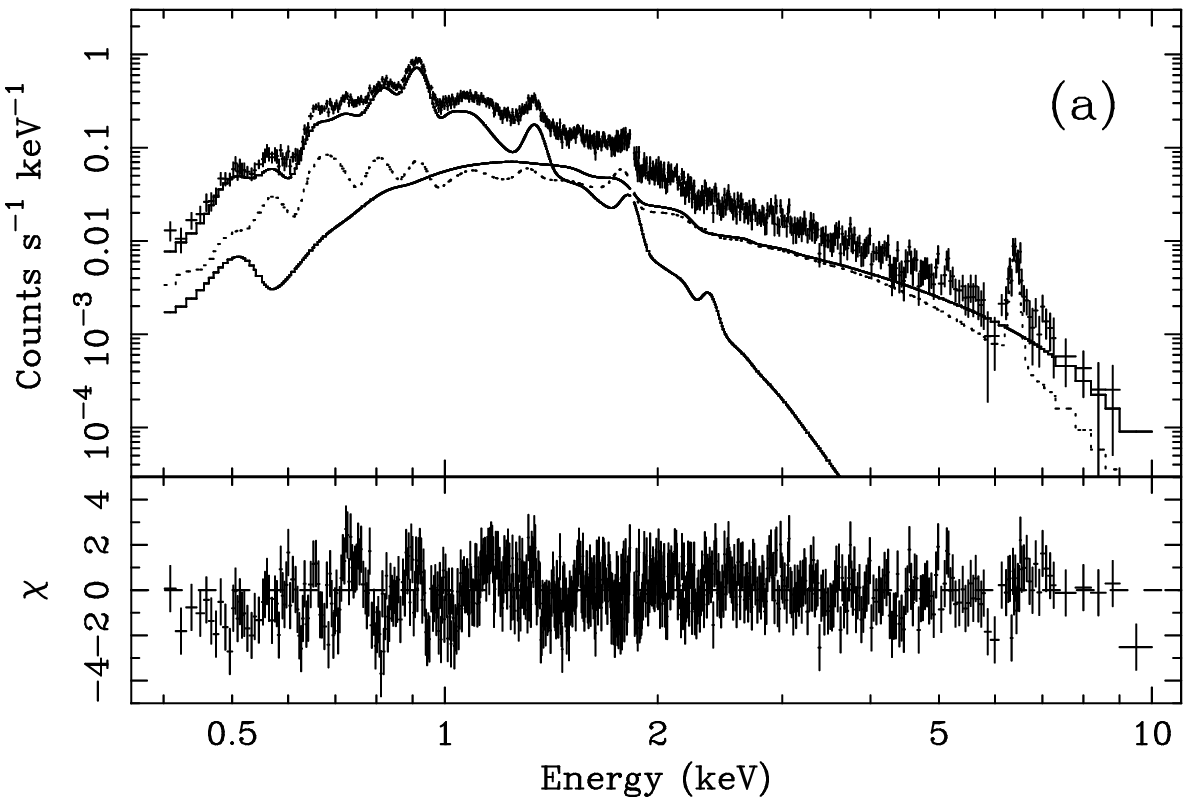}
    \FigureFile(80mm,80mm){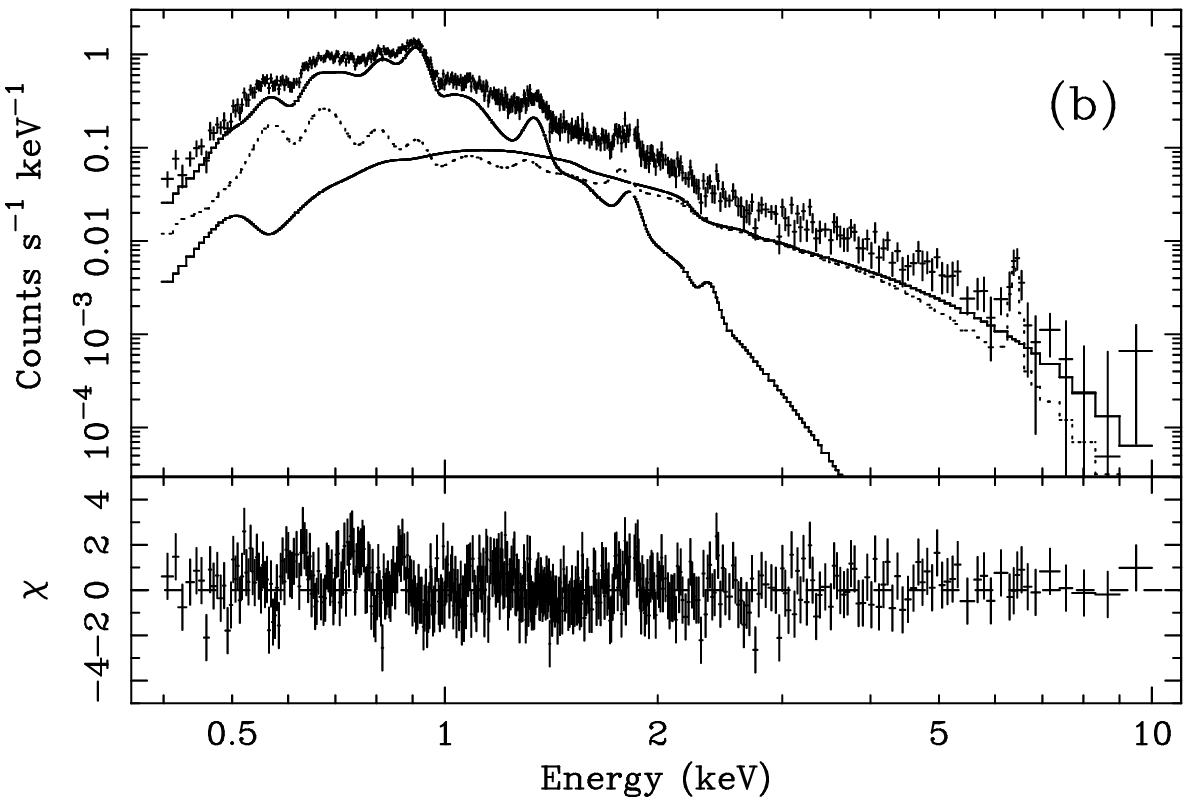}
    \FigureFile(80mm,80mm){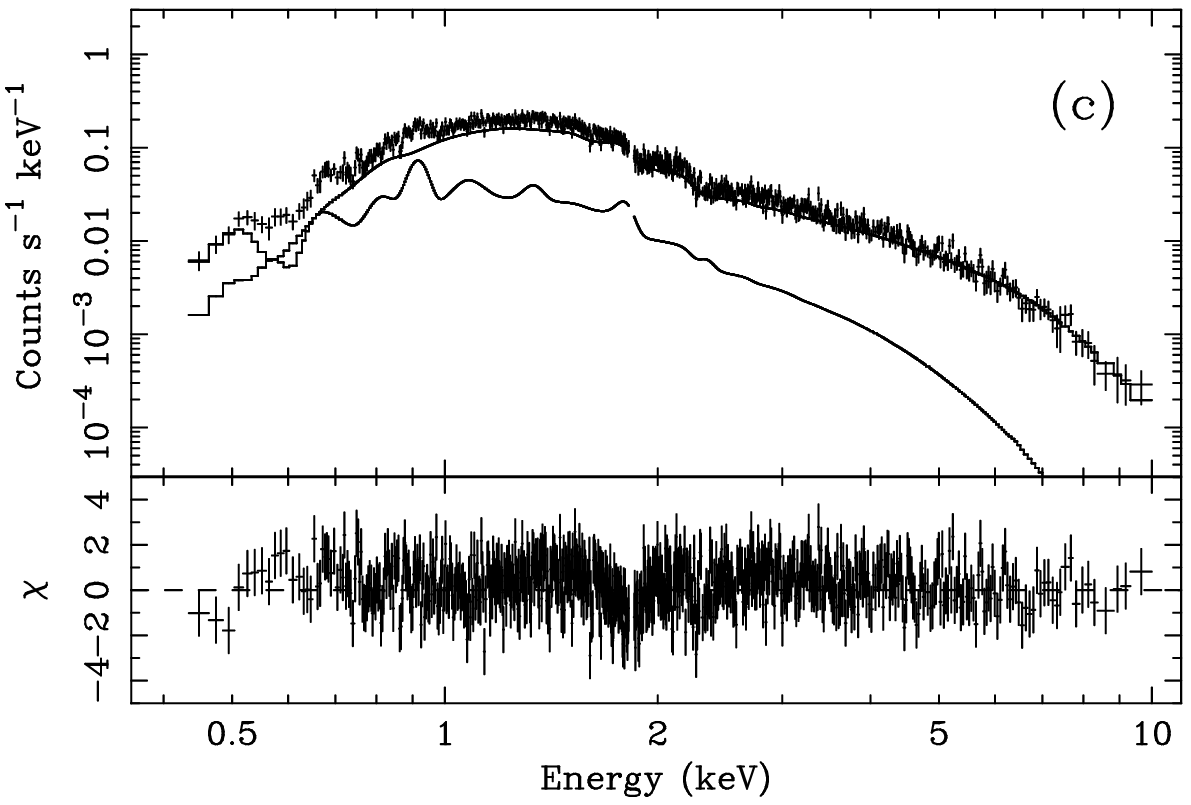}
    \FigureFile(80mm,80mm){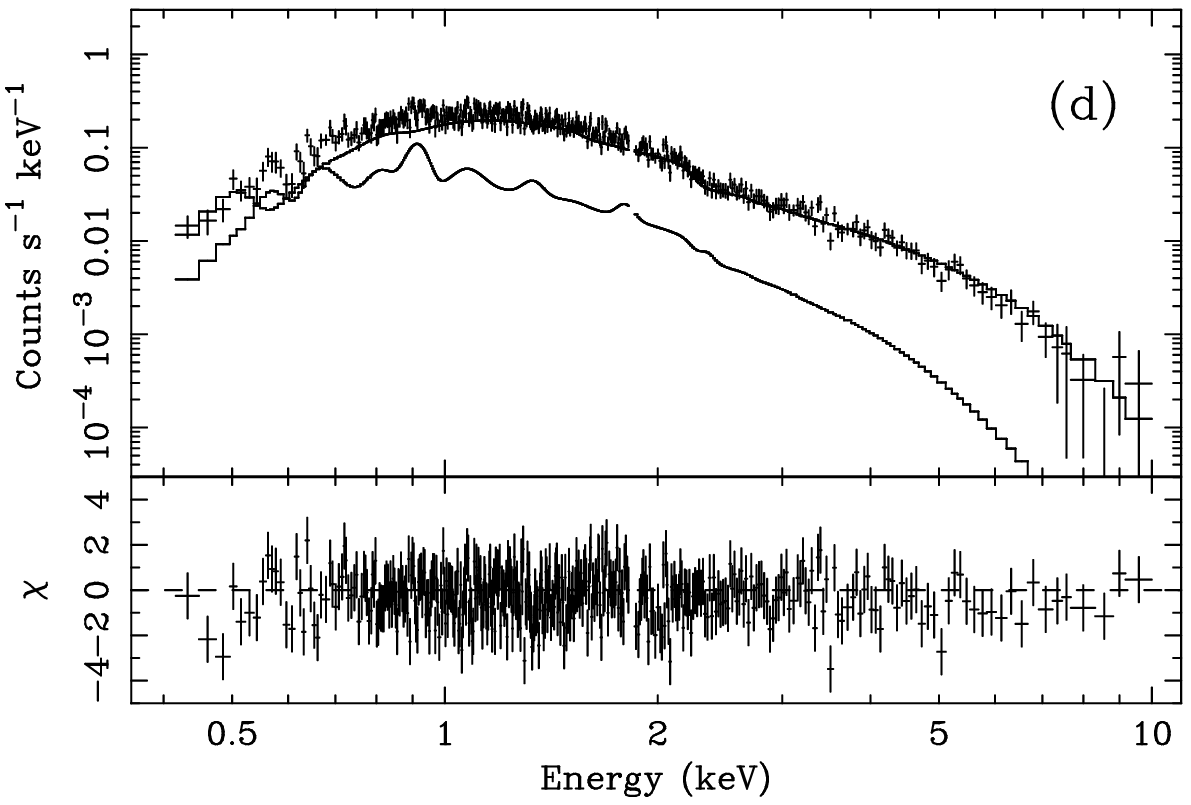}
  \end{center}
  \caption{Top: 0.4--10~keV spectra of the FI (a) and BI (b) for the East. 
    The components of the best-fit model are shown with the solid lines 
    [low-temperature plasma (VPSHOCK~1 in table~\ref{tab:full}) and power-law] 
    and dotted line  (high-temperature plasma 
    [VPSHOCK~2 in table~\ref{tab:full})]. \ 
    Bottom: The spectra of the FI (c) and BI (d) for the NE. 
    Since these spectra have no significant Fe-K line, 
    the high-temperature plasma component did not be added.}
  \label{fig:full}
\end{figure*}

\subsection{Full Band Spectrum}
\label{ssec:full}

Figure~\ref{fig:full} shows the 0.4--10~keV spectra from 
the regions of East and NE. 
The data reduction is the same as described in subsection~\ref{ssec:hard}. 
We can clearly see several emission lines below $\sim$2~keV 
in the East spectra. 
This suggests that the soft X-rays are dominantly from 
an optically thin thermal plasma. 
Therefore, we fitted the spectra from the East region 
with a thin-thermal plasma model and a power-law component. 
For the thermal plasma component, we used a VPSHOCK 
(variable abundance plane-parallel shock) model with the NEIvers 1.1 code 
(since the NEIvers 2.0 code does not include K-shell emission lines 
from low ionized Fe, we do not use this new version). 
We treated the abundances relative to solar (Anders and Grevesse~1989) 
to be free parameters, with the following elements in the parenthesis fixed to 
be the same: (C, N, and O), (Si and S), and (Ca, Fe, and Ni). 
The data in the 1.83--1.85~keV band were ignored because of 
the current calibration errors of the XIS; there is 
a small energy gap ($\sim$10~eV) at the Si K-edge that is 
not implemented in the current response function. 
Since the absolute gain of the XIS has an uncertainty 
(especially for low-energy X-rays), we allowed a small offset 
in the energy scale, independently between the FI and BI.

This 2-component model gives nice fits to the soft X-ray band 
including all the emission lines and the hard continuum emission 
with a best-fit $kT_e$ value of the thermal component of 0.38~keV 
and a power-law index of 2.9.
However it cannot reproduce the Fe-K line flux, 
with the unacceptable $\chi ^2$/d.o.f.~of 1222/1054. 
As given in subsection~\ref{ssec:hard}, the center energy of 
the Fe-K line is higher than that of Fe\emissiontype{I}, 
which suggests that the origin of the Fe-K line is due to a low 
ionized thin-thermal plasma, but is not the fluorescence of neutral Fe. 
We hence added another thin-thermal component with 
a higher temperature and a low ionized parameter. 
The abundances in this plasma were fixed to solar, except for Fe. 
Then, the best-fit $\chi ^2$/d.o.f.~was significantly reduced to be 1093/1050. 
The best-fit parameters and models are respectively 
shown in table~\ref{tab:full}, 
and in figure~\ref{fig:full} with solid and dotted lines.

We next fitted the spectra of the NE region. 
We can see emission lines in the soft X-ray band in these spectra. 
This suggests that the NE also includes a low-temperature plasma. 
However, the line profiles are too weak to obtain a reasonable fit 
with many free parameters of the low-temperature plasma. 
We therefore fixed the soft plasma parameters as those of the East region. 
Since no Fe-K$\alpha$ was found (see subsection~\ref{ssec:hard}), 
we did not include the thermal component with higher temperature. 
Thus, normalizations of the low-temperature plasma and 
power-law component are free parameters. 
To obtain the best-fit, we also fine-tuned a temperature and 
an ionization parameter in the low-temperature plasma and 
the photon index in the power-law component.
The best-fit $\chi ^2$/d.o.f.~is an acceptable value of 1138/1084. 
The best-fit parameters and spectra are given in table~\ref{tab:full} 
and figure~\ref{fig:full}. 

The thermal plasma parameters derived for the NE differ from the values 
reported by Vink et al.~(2006); we find a higher $\tau$ and a lower $kT_e$. 
Indeed, $kT_e$ and $\tau$ are correlated in the sense 
that both a higher $kT_e$ or a higher $\tau$ give a higher ionization. 
Note that Vink et al.~(2006) used a different spectral fitting code (SPEX). 
Both fits agree about the ionization state of the plasma, 
but differ concerning the relative contributions of $kT_e$ and $\tau$. 
For the NE region, the exact value for $kT_e$ is difficult to determine, 
since most of the continuum has a non-thermal origin.

\section{Discussion} 
\label{sec:discussion}

From morphology and spectral analyses, the X-rays in the northeastern 
quadrant of RCW~86 can be separated into three distinct components: 
the low temperature (VPSHOCK~1) and high temperature (VPSHOCK~2) 
thermal components both in ionization non-equilibrium, 
and the power-law component. 
We separately discuss the origin of these components, 
and finally try to give a unified picture about them.

\subsection{Low-Temperature Plasma}
\label{ssec:low}

As shown in figures~\ref{fig:image} and \ref{fig:proj}, 
the soft X-rays are dominated in the outer shell of the East region. 
The spectrum is well represented by a thin-thermal plasma with 
a low temperature of $kT_e \sim 0.3$~keV. 
Since the H$\alpha$ filament was reported to be localized 
near at the East region (Smith 1997), 
the low-temperature plasma (VPSHOCK~1) is likely to be due to 
the interstellar medium (ISM) heated by a blast wave.

The metal abundances that we obtained 
are significantly lower than the solar values. 
A sub-solar abundance is also reported in the rim region of the Cygnus Loop 
(e.g., Miyata et al.~2007), where an ISM component is dominant. 
Therefore, shocked ISM spectra may generally show such low abundances. 
In the case of RCW~86, Bocchino et al.~(2000) found 
the thermal spectra of the southeast and north rim show 
lower metal abundances than the solar values, and claimed 
that this result is due to metal depletion behind the shock. 
Since the abundances of our result are almost consistent with 
those of Bocchino et al.~(2000), the same interpretation would be accepted. 
However, we should note that the complexity of our three-component model 
makes it difficult to determine the absolute values of the abundances. 
Indeed, if we fix the Ne abundance to be solar, and fit the full band 
spectra of the East region, the best-fit can be obtained with 
an acceptable $\chi ^2$/d.o.f.~(1103/1051), and almost the same 
relative abundances to those of table~\ref{tab:full}. 
An exact separation of the low and high-temperature components and 
the power-law component is necessary for a certain determination of 
the absolute metal abundances.

The solid angle of the East ellipse in figure~\ref{fig:image} 
(= 41~arcmin$^2$) corresponds to 2.5$\times 10^{38}$~cm$^2$ at 2.8~kpc. 
Assuming the depth of the emitting region to be 8.5~pc, 
about half of the radius of the remnant, the emission volume is 
estimated to be $V = 6.6\times 10^{57}$~cm$^3$. 
Therefore, the emission measure ($EM = n_en_{\rm H}V$) of the VPSHOCK~1 
in this region corresponds to 
a proton density of $n_{\rm H} = 0.68~f_{E1}^{-0.5}$~cm$^{-3}$, 
where $f_{E1}$ is the filling factor for this component. 
Since the soft thermal emission concentrates at the outer thin rim 
in the East ellipse (see figure~\ref{fig:image}), 
the density of the low-temperature plasma is likely to be 
much larger than 0.68~cm$^{-3}$.

Similarly, the density of the VPSHOCK~1 in the NE region is estimated 
to be $n_{\rm H} = 0.26~f_{N1}^{-0.5}$~cm$^{-3}$, 
where $f_{N1}$ is the filling factor for the VPSHOCK~1 in the NE. 
Although several uncertainties remain (e.g., the filling factors, 
the temperature difference between the East and NE), 
the density of the swept-up ISM may be larger 
at the East rim than at the NE rim. 
The variation of the ionization parameter is consistent with this result, 
namely $\tau$($= n_et$) in the East is larger than that of the NE.

\subsection{Fe-K Line and High-Temperature Plasma}
\label{ssec:high}

By imaging (subsection~\ref{ssec:image}) and spectral 
(subsection~\ref{ssec:hard}) analyses, we proved that the Fe-K$\alpha$ 
emission has no correlation with the hard X-ray continuum. 
Therefore, the origin of the Fe-K$\alpha$ line is not due to 
fluorescence caused by supra-thermal electrons, nor non-thermal X-rays. 
It is located behind the blast wave (VPSHOCK~1) and is over-abundant in Fe. 
We hence propose that the origin of a high-temperature plasma (VPSHOCK~2) is 
a Fe-rich ejecta heated by a reverse shock.

The ionization parameter in VPSHOCK~2, determined from the center energy of 
Fe-K$\alpha$, is extremely low, $\tau \sim 2.3\times 10^9$~cm$^{-3}$~s. 
The electron density of the ejecta component is estimated to be 
$n_e = 0.19~f_{E2}^{-0.5}$~cm$^{-3}$, 
where $f_{E2}$ is the filling factor for the VPSHOCK~2 in the East region. 
We then estimate the elapsed time since the ejecta was 
heated by the reverse shock: 
~$t = \tau /n_e = 1.2\times 10^{10}f_{E2}^{0.5}$~sec $\lesssim 380$~yr. 
This is much less than the age of $\sim$1800~yr (from the historical record), 
and hence the ejecta must have been heated very recently.

The estimated mass of Fe is $\sim 0.07f_{E2}^{0.5}\Mo$, 
which is reasonable as a portion of the ejecta, and 
some of the Fe-ejecta may still remain in the interior of the remnant, 
which has not yet been heated by the reverse shock. 
However, we have to note the uncertainty in the mass determination. 
Although we show in subsection~\ref{ssec:full} that the spectrum of 
the East region is well reproduced by the three-component model, 
we cannot conclude this model is unique. 
We therefore, estimated the systematic error on the Fe mass 
by fitting the 3--10~keV spectrum with 
(1) only one VPSHOCK model (without a power-law), or 
(2) a pure Fe plasma and a power-law. 
In both cases, we fixed the values of $kT_e$ and $\tau$ to 1.8~keV and 
$2.3\times 10^9$~cm$^{-3}$~s, the best-fit parameters given in  
table~\ref{tab:full}. 
Then, the Fe mass was obtained to be $0.06f_{E2}^{0.5}\Mo$ and 
$0.6f_{E2}^{0.5}\Mo$ for cases of (1) and (2), respectively. 
These values can be regarded as the lower and upper limits of 
the Fe mass in the East region. 
Thus, there is an uncertainty of one order of magnitude.

\subsection{Non-Thermal Emission}
\label{ssec:non-thermal}

Since the non-thermal X-rays are well represented by 
a power-law with photon indices of $\sim 2.8$, 
synchrotron radiation is the most plausible origin. 
The synchrotron filament of the NE rim is spatially connected from 
the soft thermal rim (the East) along the outer shell of RCW~86 
(also see figure~1 of Vink et al.~2006). 
Therefore, this filament may be produced by the blast wave.

In the case of the SW rim of RCW~86, Rho et al.~(2002) and 
Ueno et al.~(2007) suggested that the non-thermal filament is not 
a blast wave, but reverse shock, because the non-thermal filament is 
inside the soft thermal emission on the SNR shell. 
Also the blast-wave velocity at the SW shell 
($\sim$800~km~s$^{-1}$: Rosado et al.~1996;
$\sim$600~km~s$^{-1}$: Ghavamian et al.~2001) is too slow to accelerate 
electrons by the standard diffusive shock acceleration theory, 
and to emit intense synchrotron X-rays. 
Since the gas density of the NE shell is $\sim$10-times lower than that of 
the SW shell (Pisarski et al.~1984), it is possible that the velocity of 
the NE blast wave is much higher than that of the present SW shell. 
Vink et al.~(2006) interpreted that the velocity of the NE non-thermal 
filament should be $\sim$2700~km~s$^{-1}$. 
This high speed can accelerate electrons to more than TeV energy 
to produce synchrotron X-rays in the blast wave shock.

\subsection{Unified Picture for All of the Components}
\label{ssec:unify}

\begin{figure}[tb]
  \begin{center}
    \FigureFile(80mm,80mm){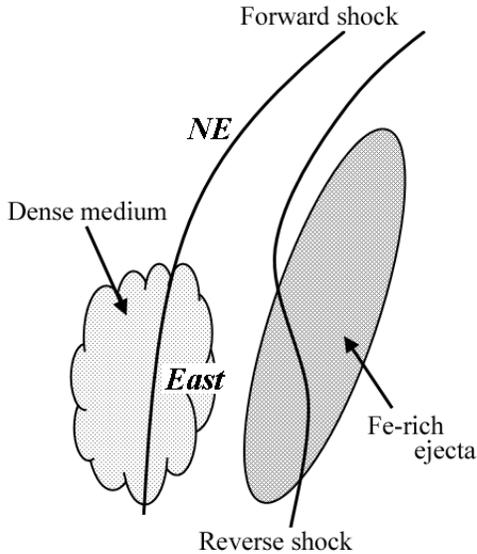}
  \end{center}
  \caption{Schematic view of the interpreted structure of 
    the RCW~86 northeast region.
  }
  \label{fig:sketch}
\end{figure}

We propose a unified picture to explain all of the results 
in figure~\ref{fig:sketch}. 
A few hundred years ago, blast waves in the East and NE rims would be 
expanding with the same high velocity. 
However the blast wave in the East collided with a dense medium 
very recently, and the forward shock decelerated rapidly. 
At that same time, the reverse shock behind the East rim began to move 
inward to the interior of the remnant, and it heated the Fe-rich ejecta. 
Since RCW~86 may be a remnant in the OB association (Westerlund 1969), 
a candidate for the ``dense medium'' is either a cavity wall surrounding 
the remnant, as suggested by Vink et al.~(1997), or a molecular cloud. 
We found in the NANTEN results (Matsunaga et al.~2001) evidence of a cloud 
just at the position of the East rim of RCW~86 ($l$=315.7, $b$=--2.4). 
We strongly propose a deeper radio observation to reveal 
whether the shell of the remnant is truly interacting with the cloud.

The forward shock in the NE rim, on the other hand, is still expanding 
in a tenuous region, and hence keeps a high shock velocity. 
The reverse shock behind the NE may also expand with high velocity, 
and hence has not yet reached the Fe-rich ejecta layers.
This may explain the absence of Fe-K emission behind the NE region.

\section{Summary}
\label{sec:summary}

We have analyzed Suzaku/XIS data obtained from 
the northeast quadrant of RCW~86. 
The results and interpretations are summarized as follows:

\begin{enumerate}

\item  The Fe-K$\alpha$ emission is enhanced at the inner region 
  from the soft thermal rim (the East region), and 
  has no correlation with the hard X-ray non-thermal filament (the NE region). 

\item  The centroid of the Fe-K$\alpha$ line (6404--6444~eV) and 
  the lack of Fe-L lines constrain the Fe ionization state to be 
  approximately between Ar-like (Fe\emissiontype{IX}) and 
  Ne-like (Fe\emissiontype{XVII}). 

\item  The spectra extracted from the soft X-rays and 
  Fe-K$\alpha$ enhanced region (the East) are well fitted with 
  two thin-thermal plasmas, which have different temperature of 
  $kT_e\sim 0.3$~keV and $\sim 1.8$~keV 
  and one power-law with a photon index of $\Gamma \sim 2.9$. 

\item  The spectra extracted from the hard X-ray filament (the NE) 
  can be represented by a lower temperature plasma and a power-law. 

\item  The lower temperature plasma has sub-solar metal abundances. 
  The origin of this component would be a blast-shocked ISM. 

\item  The higher temperature plasma includes an over-solar abundant iron, 
  and has an extremely low plasma age of 
  $\tau \sim 2.3\times 10^9$~cm$^{-3}$~s. 
  Hence it produces nearly 6.4~keV K$\alpha$ lines. 
  This component may be Fe-rich ejecta heated by reverse shock 
  very recently. 

\item  The power-law component, which is the brightest in the NE region, 
  can be regarded as synchrotron emission. 

\item  The shock front at the East rim would recently have collided 
  with dense ISM, after which the blast wave decelerated rapidly, and 
  the reverse shock began to move toward the center of the remnant. 

\item  The blast wave at the NE rim may still be expanding in 
  a tenuous region. Therefore, it keeps a high shock velocity for 
  efficient particle acceleration. 

\end{enumerate}

\bigskip

The authors thank all members of the Suzaku team. H.~Y., H.~N., and A.~B. are 
supported by JSPS Research Fellowship for Young Scientists. 
This work is supported by the Grant-in-Aid for the 21st Century COE 
"Center for Diversity and Universality in Physics" from the Ministry of 
Education, Culture, Sports, Science and Technology (MEXT) of Japan, 
and by Grants-in-Aid for Scientific Research of the Japanese Ministry of 
Education, Culture, Sports, Science, and Technology (K.~K., R.~Y., and A.~K.).



\begin{thebibliography}{}

\bibitem[Anders \& Grevesse(1989)]{1989GeCoA..53..197A} Anders, E., \& 
Grevesse, N.\ 1989, \gca, 53, 197 

\bibitem[Bamba et al.(2000)]{2000PASJ...52.1157B} Bamba, A., Koyama, K., \& 
Tomida, H.\ 2000, \pasj, 52, 1157 

\bibitem[Bocchino et al.(2000)]{2000A&A...360..671B} Bocchino, F., Vink, 
J., Favata, F., Maggio, A., \& Sciortino, S.\ 2000, \aap, 360, 671 

\bibitem[Borkowski et al.(2001)]{2001ApJ...550..334B} Borkowski, K.~J., 
Rho, J., Reynolds, S.~P., \& Dyer, K.~K.\ 2001, \apj, 550, 334 

\bibitem[Ghavamian et al.(2001)]{2001ApJ...547..995G} Ghavamian, P., 
Raymond, J., Smith, R.~C., \& Hartigan, P.\ 2001, \apj, 547, 995 

\bibitem[Ishisaki et al.(2007)]{2007PASJ...59S.113I} Ishisaki, Y., et al.\ 
2007, \pasj, 59, S113 

\bibitem[Koyama et al.(1995)]{1995Natur.378..255K} Koyama, K., Petre, R., 
Gotthelf, E.~V., Hwang, U., Matsuura, M., Ozaki, M., \& Holt, S.~S.\ 1995, 
\nat, 378, 255 

\bibitem[Koyama et al.(2007)]{2007PASJ...59S..23K} Koyama, K., et al.\ 
2007, \pasj, 59, S23 

\bibitem[Matsunaga et al.(2001)]{2001PASJ...53.1003M} Matsunaga, K., 
Mizuno, N., Moriguchi, Y., Onishi, T., Mizuno, A., \& Fukui, Y.\ 2001, 
\pasj, 53, 1003 

\bibitem[Mitsuda et al.(2007)]{2007PASJ...59S...1M} Mitsuda, K., et al.\ 
2007, \pasj, 59, S1 

\bibitem[Miyata et al.(2007)]{2007PASJ...59S.163M} Miyata, E., Katsuda, S., 
Tsunemi, H., Hughes, J.~P., Kokubun, M., \& Porter, F.~S.\ 2007, \pasj, 59, 
S163 

\bibitem[Pisarski et al.(1984)]{1984ApJ...277..710P} Pisarski, R.~L., 
Helfand, D.~J., \& Kahn, S.~M.\ 1984, \apj, 277, 710 

\bibitem[Rho et al.(2002)]{2002ApJ...581.1116R} Rho, J., Dyer, K.~K., 
Borkowski, K.~J., \& Reynolds, S.~P.\ 2002, \apj, 581, 1116 

\bibitem[Rosado et al.(1996)]{1996A&A...315..243R} Rosado, M., 
Ambrocio-Cruz, P., Le Coarer, E., \& Marcelin, M.\ 1996, \aap, 315, 243 

\bibitem[Serlemitsos et al.(2007)]{2007PASJ...59S...9S} Serlemitsos, P.~J., 
et al.\ 2007, \pasj, 59, S9 

\bibitem[Smith(1997)]{1997AJ....114.2664S} Smith, R.~C.\ 1997, \aj, 114, 
2664 

\bibitem[Stepehson(2002)]{2002Histrical SNR} Stephenson, F. R., \& Green, D. A.
\ 2002, Historical Supernovae and their Remnants 
(Oxford: Oxford University Press) 

\bibitem[Tomida et al.(1999)]{1999AN....320..342T} Tomida, H., Koyama, K., 
\& Yamauchi, S.\ 1999, Astron. Nachr. 320, 342 

\bibitem[Ueno et al.(2007)]{2007PASJ...59S.171U} Ueno, M., et al.\ 2007, 
\pasj, 59, S171 

\bibitem[Vink et al.(1997)]{1997A&A...328..628V} Vink, J., Kaastra, J.~S., 
\& Bleeker, J.~A.~M.\ 1997, \aap, 328, 628 

\bibitem[Vink et al.(2006)]{2006ApJ...648L..33V} Vink, J., Bleeker, J., van 
der Heyden, K., Bykov, A., Bamba, A., \& Yamazaki, R.\ 2006, \apj, 648, 
L33 

\bibitem[Westerlund(1969)]{1969AJ.....74..879W} Westerlund, B.~E.\ 1969, 
\aj, 74, 879 

\end{thebibliography}
\end{document}